%% file: paper.tex
\documentclass[12pt,preprint]{aastex}

\newcommand{\myemail}{yuan@noao.edu; qirongwy@public1.ptt.js.cn}
\newcommand{\wavwav}[1]{{$\lambda\lambda$}#1}
\newcommand{\kms}{km~s$^{-1}$}

\shorttitle{Associated absorption lines in 3C~351} \shortauthors{Yuan et al.}

\received{2002 January 31}
\begin{document}

\title{Associated Absorption Lines in the Radio-Loud Quasar 3C~351:
Far-Ultraviolet Echelle Spectroscopy from the Hubble Space Telescope}

\author{ Q. Y{\small UAN}\altaffilmark{1,2} R. F. G{\small
REEN}\altaffilmark{2}, M. B{\small ROTHERTON}\altaffilmark{2}, T. M. T{\small
RIPP}\altaffilmark{3}, M. E. K{\small AISER}\altaffilmark{4}, G. A. K{\small
RISS}\altaffilmark{5}}

\altaffiltext{1}{Department of Physics, Nanjing Normal University,
                NingHai Road 122, Nanjing 210097, China; \myemail}

\altaffiltext{2}{Kitt Peak National Observatory, National Optical Astronomy
                Observatory, 950 N. Cherry Ave.,
                P.O. Box 26732, Tucson, AZ 85726-6732;
                green,mbrother@noao.edu}

\altaffiltext{3}{Princeton University Observatory, Princeton, NJ 08544}

\altaffiltext{4}{Department of Physics and Astronomy, Johns Hopkins
University, Baltimore, MD 21218}

\altaffiltext{5}{Space Telescope Science Institute, 3700 San Martin Drive,
               Baltimore, MD 21218}

\begin{abstract}

As one of the most luminous radio-loud quasars showing intrinsic ultraviolet
(UV) and X-ray absorption, 3C~351 provides a laboratory for studying
the kinematics and physical conditions of such ionized absorbers. We present
an analysis of the intrinsic absorption lines in the high-resolution ($\sim$ 7
\kms) far-UV spectrum which was obtained from observations
with the {\em Space Telescope Imaging Spectrograph} (STIS) on board
the {\em Hubble Space Telescope} ($HST$). The spectrum spans wavelengths from
1150 \AA ~to 1710 \AA, and shows strong emission lines from \ion{O}{6} and
Ly$\alpha$. Associated absorption lines are present on the blue wings of the
high-ionization emission doublets \ion{O}{6} \wavwav{1032,1038} and \ion{N}{5}
\wavwav{1238,1242}, as well as the Lyman lines through Ly$\epsilon$. These
intrinsic absorption features are resolved into several distinct kinematic
components, covering rest-frame velocities from $-$40 to $-$2800 \kms, with
respect to the systemic redshift of $z_{em}=0.3721$. For the majority of these
absorption line regions, strong evidence of partial covering of both the
background continuum source and the broad emission line region (BELR) is found,
which supports the intrinsic absorption origin and rules out the possibility
that the absorption arises in some associated cluster of galaxies. The
relationship between the far-UV absorbers and X-ray `warm' absorbers are
studied with the assistance of photoionization models. Most of the UV
associated absorption components have low values of the ionization parameter
and total hydrogen column densities, which is inconsistent with previous claims
that the UV and X-ray absorption arises in the same material. Analysis of these
components supports a picture with a wide range of ionization parameters,
temperatures, and column densities in AGN outflows.

\end{abstract}



\section{INTRODUCTION}

It has been recently appreciated that the ionized (or so-called ``warm'') gas
that produces intrinsic absorption in the X-ray spectra of low-luminosity
Active Galactic Nuclei (AGNs) (e.g., Seyfert 1 galaxies) is an important,
newly recognized component of their near-nuclear structure. Recent X-ray
spectroscopic observations indicate that more than half of Seyfert 1 galaxies
show K-shell absorption edges of warm oxygen (\ion{O}{7} and \ion{O}{8})
characteristic of photoionized gas (Reynolds 1997; George et al. 1998).
Moreover, a similar fraction show intrinsic absorption associated with their
active nuclei in their ultraviolet (UV) spectra, and there appears to be a
one-to-one correspondence between objects showing X-ray and UV absorption,
suggesting that these two phenomena are related \citep{crenshaw99}.

The amount and presence of intrinsic absorption are found to depend upon either
the luminosity or radio properties of AGNs: the narrow ``associated'' ($z_{ab}
\approx z_{em}$) UV and X-ray ``warm'' absorbers are rare in high-luminosity
and radio-loud quasars \citep{nicastro99}. Among high-luminosity quasars,
intrinsic broad absorption lines (BALs) spanning up to tens of thousands of
\kms are seen in about 10-20\% of quasars, with a decrease in outflow velocity
in the radio-loud quasars \citep{becker01}. BAL quasars are also deficient in
X-rays (Green et al. 2001; Gallagher et al. 2002). The luminous radio-loud
quasars with narrow absorption lines (NALs) appear to bridge the Seyfert warm
absorbers and the BAL quasars, and may possess different physical
conditions. This category includes two well studied radio-loud quasars,
namely, 3C~351 \citep{fiore93}, and 3C~212 \citep{mathur94},  which are found
to have ionized absorbers. In the current paper we pursue more detailed
investigations of the the associated UV absorbers in 3C~351 based on a
high-resolution far-UV spectrum.

Observationally, the radio-loud quasar 3C~351 ($z_{em}=0.3721$; Marziani et al.
1996) is very lobe-dominated with only 0.65\% of the flux density from the
compact core at 6 cm \citep{kellermann89}, showing two opposite jets of
similar angular extent. This FR II source has a steep radio spectrum and
a low ratio of radio core luminosity at 5 GHz to optical V-band continuum
luminosity ($log R_V \sim$ 0.73; Brotherton 1996), indicating an
edge-on geometry of the central engine. Additionally,
3C~351 is also rather X-ray-``quiet'' with a factor of $\sim$ 5 lower X-ray
flux than the average radio-loud quasar \citep{wilkes94}. Its effective
optical-to-X-ray slope $\alpha_{ox}$ is 1.55 \citep{tananbaum86}, as compared
to $\alpha_{ox}=1.3$ for an average radio-loud quasar. Even though 3C 351 is
X-ray quiet, among radio-loud quasars known to have ionized absorption, it
is one of the most X-ray luminous($L_{0.1-2keV}=2.3 \times
10^{45}$ ergs s$^{-1}$; Fiore et al. 1993), and therefore it is an excellent
test case for studying the kinematics and physical conditions of the absorbing
outflow associated with the nucleus of a radio-loud quasar.

The {\em ROSAT} Position Sensitive Proportional Counter (PSPC) observations
show that the dominant feature in its X-ray spectrum is a strong \ion{O}{7}
and/or \ion{O}{8} absorption edge. A factor of 1.7 change in flux was
found between observations in 1991 October and 1993 August, which stimulated a
test of ionization models for the warm absorber in 3C~351, under the simplest
photoionization equilibrium \citep{nicastro99}. Moreover, as a part of the
Quasar Absorption Lines Key project \citep{bahcall93}, an
intermediate-resolution (R $\sim$ 1300) UV spectrum of 3C~351 was taken with
the Faint Object Spectrograph (FOS) of the {\em HST} . As a result, a strong
absorption system at $z=0.3646$ was found in the high-ionization doublets of
\ion{O}{6}, \ion{N}{5}, and \ion{C}{4}, as well as in the Lyman series from
Ly$\alpha$ to Ly$\delta$. It was suggested by the authors that these UV
absorption lines are due to an associated cluster of galaxies. However, no
compelling evidence of the cluster was found in follow-up studies
\citep{ellingson94,lanzetta95,lebrun96}. In this paper we provide profound
evidence that this associated absorption system is an intrinsic absorber
located close to the QSO nucleus.

To what extent do the UV and X-ray absorbers arise in the same regions?
This is an important question for understanding the structure of quasar
nuclei. In the case of 3C~351, \citet{mathur94} proposed a simple one-zone
photoionization model and claimed that the UV absorption features are due to
the same material detected in the X-ray observations. To answer this question,
quasi-simultaneous high-resolution UV and X-ray spectra of intrinsic lines
should be taken. We notice that the UV absorptions presented in previous
studies are typically comprised of multiple kinematic components
\citep{crenshaw99,kriss00}. High-resolution spectroscopic observations allow
accurate determination of absorption width and resolve multiple components,
which is the key in determining the column density and velocity structure along
the line of sight.

The Space Telescope Imaging Spectrograph (STIS) was installed during the second
servicing mission of {\it HST}, and the high-resolution {\em E140M} and
{\em E230M} echelle modes of this instrument are ideally suited for the
study of associated UV absorbers. In this paper we present a high-resolution,
far-UV STIS spectrum of 3C~351 covering the \ion{O}{6} \wavwav{1032,1038},
\ion{N}{5} \wavwav{1238,1242} and the Lyman series (\S 2).  We focus on
resolving the kinematic components in the associated absorption (\S 3), and
interpret the spectral features with the assistance of photoionization models
(\S 4). Then, we discuss our findings together with previous results for a
better understanding of the ionized outflow of 3C~351 (\S 5). Finally, a
summary is given (\S 6).

\section{OBSERVATIONS}

High-resolution echelle spectra of 3C351 were obtained with STIS on four
occasions between 1999 June 27 and 2000 July 25. A brief log of the
observations is provided in Table~\ref{obslog} including the date, total
exposure time, and {\it HST} archive identification codes for each
observation. Individual exposure durations ranged from 2230 to 2500 seconds,
and wavelength calibration exposures were obtained between the individual
observations of the quasar. All of the observations employed the
intermediate-resolution {\em E140M} echelle mode of STIS with the $0\farcs 2
\times 0\farcs 06$ slit; this mode provides a resolution of 7 \kms (FWHM) with
nearly complete wavelength coverage from 1150 $-$ 1710 \AA\ (There are only
four small gaps between orders at $\lambda >$ 1634 \AA ). A complete
description of the design of STIS has been provided by \citet{woodgate98}, and
\citet{kimble98} assess the on-orbit performance of the instrument.

The data were reduced at the Goddard Space Flight Center with the software
developed by the STIS Investigation Definition Team (IDT). Individual exposures
were extracted following standard procedures for flatfielding and flux and
wavelength calibration. Scattered light was removed using the
scattered light correction developed by the STIS IDT, which accounts for
echelle scatter as well as other sources of scattered light. Then the
individual exposures were coadded weighted by their inverse variances averaged
over a large region of one of the central orders. Finally, overlapping regions
of adjacent orders were coadded again weighted by the inverse variance in each
pixel, but in this case the variance vector was smoothed with a 5-pixel boxcar
before being used for determination of the weighting so that noise
fluctuations do not lead to inappropriate over or under-weighting. We
initially reduced the data from the four observation dates separately.
Careful inspection of the resulting four individual spectra reveals no
significant variability. Consequently, we elected to coadd all of the data
to obtain a single, high signal-to-noise spectrum. The total exposure time
of the final spectrum is 78,198 seconds.

The overall appearance of the final spectrum of 3C351 is shown in Figure 1
(nine-pixel smoothing is used for this display, and the 1$\sigma$ flux
uncertainty is given). The broad \ion{O}{6} and Ly$\alpha$ emission lines are
readily apparent. In this paper, we are mainly interested in the strong
associated \ion{O}{6} and Ly$\alpha$ absorption lines visible on the blue side
of the emission lines, as well as the associated \ion{N}{5} lines near 1700
\AA. The spike at 1216 \AA\ is the geocoronal Ly$\alpha$ emission line, and the
broad trough at 1216 is the damped Ly$\alpha$ absorption line due to \ion{H}{1}
in the Milky Way ISM. Some of the absorption lines in this spectrum are due to
the low-redshift Ly$\alpha$ forest; investigation of these lines is valuable
for understanding the nature and properties of the intergalactic medium in the
nearby universe (e.g., Dav\'{e} \& Tripp 2001).

\section{SPECTRAL MEASUREMENTS}

\subsection{Continuum and Emission Features}

We used the $IRAF$ task {\sf specfit} (ver8.6) \citep{kriss94} to fit the STIS
far-UV spectrum. The continuum was modeled as a power law, and only a
Galactic extinction correction of $A_B = 0.097$ \citep{schlegel98} was
applied by using the reddening curve of \citet{cardelli89} with $R_V=3.1$.
The power-law index obtained is $\alpha = 0.0 \pm 0.029$, where $F_{\lambda}
\propto {\lambda}^{-\alpha}$. The observed flux in the continuum at 1500~\AA
~is $\sim$ 1.5 $\times$ 10$^{-14}$ erg s$^{-1}$ cm$^{-2}$ \AA$^{-1}$.
The doublet \ion{O}{6} \wavwav{1032,1038}
emission features are each fitted with a double Gaussian profile: a broader
(FWHM $\sim$ 21,000 \kms) component with a redshift of 0.3796 and a narrower
(FWHM $\sim$ 2400 \kms) component at $z = 0.3698$. The main fraction of the
recombination emission originates in clouds in the broad emission-line region
(BELR) that are optically thick in the ionizing continuum. The doublet flux
ratio is found to be about 1:1 for both the broader and narrower components of
the broad lines, which is reasonable for the optically thick physical
conditions of the BELR \citep{hutchings01}. The doublet \ion{N}{5}
\wavwav{1238,1242} is located at the red edge of the STIS spectrum and separate
emission-line components were not required in our fitting model.

For the strong Ly$\alpha$ emission profile, two Gaussians are used to give a
good fit, one broad (FWHM $\sim$ 15,000 \kms) and one narrow (FWHM $\sim$
3600 \kms). We also searched for emission lines from other Lyman lines, but
if any are present they are very broad and weak (Our formal result finds only a
broad emission component of Ly$\gamma$). For a clear physical picture of
dynamics of the broad and narrow components in the BELR, we adopt the same
rest-frame velocities for Ly$\alpha$ as those detected in the \ion{O}{6}
doublet during the fitting of emission-line profiles. The radial velocities for
the broad and narrow emission-line components with respect to the systemic
redshift, $z_{em}=0.3721\pm 0.0003$ (2$\sigma$), obtained by \citet{marziani96}
from the analyses of $FOS$ UV spectra and ground-based optical spectroscopy
\citep{boroson92}, are 1639 and -450 \kms, respectively. The emission fluxes
and observed central wavelengths are listed in Table~\ref{emline}.

\subsection{Associated Absorption Components}

The high resolution and good S/N of this far-UV spectrum makes it possible to
resolve unambiguously intrinsic absorption components in the \ion{O}{6},
\ion{N}{5} and Lyman lines. In order to show all the distinct components of
associated absorption, we present the blue wings of the high-ionization
doublets \ion{O}{6} and \ion{N}{5} in Fig. 2(a), and a similar diagram for
the Lyman series in Fig. 2(b), where the fluxes are plotted as a function of
the rest-frame velocity. There are in total 15 components detected in the
associated absorption features (alphabetically, A -- O), among which component
G, with a relative velocity of -1699 \kms (i.e., $z=0.3644$), is the most
striking one of either the high-excitation doublets or the Lyman series.
An unusually strong metal line absorption system with a very close redshift
($z=0.3646$) to component G was detected by \citet{bahcall93} in the
intermediate-resolution ($R \sim 1300$) FOS spectroscopy, and it contains
contributions from many of the components we resolve at 7 \kms resolution. The
FOS spectrum covers a wider wavelength range from 1180\AA ~to 3270 \AA, and
this strong metal absorption system is also seen in the \ion{C}{4} doublet
\wavwav{1548,1550} at 2112.8 and 2116.6 \AA, respectively. The STIS spectrum
has such good resolution that component G in Ly$\epsilon$ is unblended with the
Ly$\gamma$ absorption from the intervening system at $z=0.3175$, and even the
weak narrow associated absorption components, such as components K, M, N, and
O, can be clearly resolved. We notice that components A (at $z=0.3719$) and G
appear in the Lyman series, \ion{O}{6}, and \ion{N}{5}. The very narrow
intrinsic components K, N and O are found only in \ion{O}{6}, and components B
and J can be seen only in the Lyman series.

Additionally, the intergalactic metal-line absorption system at $z=0.2210$
detected by \citet{boisse92} and \citet{bahcall93} is confirmed by the STIS
spectrum. The line \ion{Si}{4} 1393\AA ~of this intergalactic system
appears in the domain (at 1701.8\AA) of the \ion{N}{5} doublet, and therefore
is excluded in our measurement.

\subsubsection{Associated Absorption Lines in the Doublets of \ion{O}{6},
\ion{N}{5} and the Covering Factors}

The intrinsic absorbers near the QSO might cover only part of the continuum or
broad-line emitting sources along our line of sight \citep{hamann97}, and the
covering factors could differ between ions and vary with velocity across the
line profiles. The column density will be underestimated if this effect is not
accounted for. In the case of single absorption lines (e.g., one
component of Ly$\alpha$ absorption), we can derive a lower limit to the
covering factor, $C_f$, along the line of sight from the residual flux, $I_0$,
in the core of the normalized absorption line at a particular radial velocity
by $ C_f \geqslant 1 - I_0$.
For component B, the complete absorption in Ly$\alpha$ shows its covering
factor is nearly 100\% (i.e.,$I_0 \approx 0$) (see Fig. 4). In practice, the
actual value of the covering factor for a component can be determined only from
a doublet, assuming both absorption lines of the doublet are unblended with
other lines \citep{crenshaw99}. For the doublets \ion{O}{6} and \ion{N}{5} in
the STIS far-UV spectrum, the expected ratio of the optical depths of the
doublet is $\sim$ 2.0, and the covering factor $C_f$ can be derived by
\begin{eqnarray}
C_f & = & \left\{
\begin{array}{lll}
 \frac{I_1^2-2I_1+1}{I_2-2I_1+1}
        & {\rm for} & I_1>I_2\geqslant I_1^2, \\
1 & {\rm for} & I_2<I_1^2, \\
1 - I_1 & {\rm for} & I_2\geqslant I_1,
\end{array} \right.
\end{eqnarray}
where $I_1$ and $I_2$ are the normalized intensities in the weaker (e.g.,
\ion{O}{6} $\lambda$1037.6) and stronger (e.g., \ion{O}{6}
$\lambda$1031.9) line troughs, respectively. In most cases, the value
of $I_2$ is between $I_1^2$ and $I_1$. If $I_2$ is outside of this range,
it should be due to measurement uncertainties. The corresponding line
optical depths of this doublet, ($\tau_1, \tau_2$), are
\begin{eqnarray}
\begin{array}{lllll}
\tau_2 & = & 2\,\tau_1 & = & 2\,ln\left( \frac{C_f}{I_1+C_f-1} \right).\\
\end{array}
\end{eqnarray}

To determine the covering factor of each component, we choose the
high-ionization doublet \ion{O}{6} to start our fitting because it is
continuous and has a higher signal-to-noise ratio as compared with the
\ion{N}{5} doublet. The absorption lines are treated as Gaussian profiles in
optical depth $\tau(v_r)$, and the ratio of optical depths for this doublet is
fixed at 2:1, assuming that the \ion{O}{6} and \ion{N}{5} absorption lines are
fully resolved. Most of the broad and smooth components appear to be well
resolved in the STIS spectrum. However, there are also narrow components that
could, in principle, contain unresolved absorption. Theoretically, only
components with temperatures less than 17,000 K would be still unresolved at 7
\kms resolution.

Component A of Ly$\beta$ is blended with the broad absorption component H of
\ion{O}{6}, and its width and location are well constrained by its
corresponding absorption profiles in \ion{O}{6} and Ly$\alpha$. Since the width
of component G is constrained by the corresponding spectral profiles of
Ly$\beta$ and Ly$\gamma$, the broadest absorption troughs in the \ion{O}{6} and
Ly$\alpha$ lines around component G cannot be fitted by a single component, so
components F and H were introduced. The associated absorption features in
\ion{O}{6} were carefully deblended with 13 components (namely, A,C $-$ I,K $-$
O), and the resolving process is unambiguous. Fig. 3 shows the best fit to the
\ion{O}{6} emission and absorption features. A strong narrow absorption is
found at 1421.5\AA, just between components A and C, and the possibility of
being associated with 3C~351 can be ruled out. If it were an \ion{O}{6}
$\lambda1038$ absorption line, a stronger line of \ion{O}{6} $\lambda1032$
should appear at 1413.3\AA\ (see Fig 2a). Absence of the corresponding doublet
absorption indicates that this feature is probably Lyman $\alpha$ in the
intergalactic medium at $z=0.1693$. Additionally, two narrow absorption lines
resulting from intervening Ly$\alpha$ clouds can be found in the \ion{O}{6}
region at 1413.938 and 1420.226 \AA.

The STIS far-UV spectrum of the \ion{N}{5} doublet at $\sim$ 1700 \AA\ has
poorer S/N, as it appears on the red edge of the effective wavelength
range. As seen in Fig. 2, only the strong associated components can be
detected. We adopt the same values of rest-frame velocity, width, and
covering factor for each component in \ion{N}{5} as in \ion{O}{6} because both of
these two species probably co-exist in the same high-ionization material.
We achieved a good (in the $\chi^2$-sense) fit of the \ion{N}{5}
absorption features.

For fully resolved lines that have Gaussian distributions of optical depth,
the column densities $N_{ion}$ (in cm$^{-2}$) can be derived by integrating
the line optical depths across the absorption profiles by
\begin{equation}
\begin{array}{lll}
N_{ion}& =& \frac{m_ec}{\pi e^2\lambda_0 f} \int \tau (v_r) dv_r
   = \left(\frac{m_ec}{\pi e^2}\right) \,\, \left(\frac{\pi}{4ln2}
     \right) ^{1/2} \,\, \left(\frac{\tau_0 W}{\lambda_0 f}\right)
     \\
  &=& 4.0108 \times 10^{14} \times
    \left(\frac{\tau_0 W}{\lambda_0 f}\right) \,\,\, ({\rm cm}^{-2}) ,
\end{array}
\end{equation}
where 
$\lambda_0$ and $f$ are the transition wavelength (in \AA) and oscillator
strength, which are cataloged by \citet{morton91}, $\tau_0$ and $W$ are
central optical depth and width (i.e., FWHM in \kms) of the absorption line
\citep{savage91,arav01}. The function `ptauabs' in the latest version of
the {\sf specfit} task is used to parameterize absorption with partial
covering of the continuum and BEL sources.

Table~\ref{o6n5} presents the measurements of
the associated absorption lines in these two doublets. We give the column
density, line width (FWHM), rest-frame velocity, and covering factor for each
component. The broadest component, E, is found to have the
smallest covering factor in the \ion{O}{6} and \ion{N}{5} doublets, and its
column densities are even larger than those of component G. We emphasize
that the column density $N_{ion}$ and central optical depth
$\tau(v_r)$ depend heavily upon the covering factor, as shown in equations (2)
and (3). It is well known that spectral resolution is crucial for accurate
measurement of the associated absorption components in the doublets. Previous
studies which are based on low-resolution (R $\sim$ 1000) spectroscopic
observations may have severe biases in column density determinations, since the
blended complex of components in doublets in lower resolution spectra cannot
be expected to result in any useful information about the covering factors.

\subsubsection{Associated Absorption Lines in the Lyman series}

Absorption features of the associated systems can be found in the Lyman
lines, including all the major components (i.e., FWHM $>$ 20 \kms) that appear in
the high-ionization \ion{O}{6} doublets. For these components we adopt
the same values of the rest-frame velocity and width as those derived from the
\ion{O}{6} fitting. The covering factors of absorption components in
principle can be variable between different lines of the Lyman series since
the BELR may be uncovered and each transition could therefore cover a
different fraction of the total continuum plus emission line flux.
However, permitting this extra freedom in the fitting could compromise the
consistency of the results and so we initially set all lines to have the
same covering factor. Some strong absorption components are found to have
cores that are much deeper than the continuum flux levels (e.g., components A
and B in Ly$\alpha$), indicating that the regions responsible for the
associated absorption lines lie completely outside of the BELR.

We fit the higher order Lyman lines (Ly$\beta$, Ly$\gamma$, Ly$\delta$
and Ly$\epsilon$) simultaneously with Ly$\alpha$ since each is determined by the
same neutral hydrogen column density and known oscillator strengths. We notice that
there are strong emission features in Lyman $\alpha$ and \ion{O}{6}, whereas
there are no obvious emission features in the higher order Lyman lines. Keeping
in mind that the BELR may influence the covering factors for the associated
Ly$\alpha$ absorbers, we force Ly$\alpha$ to have the minimum covering factor
smaller than or equal to those of the higher order Lyman lines. In practice, we
initially set the same values of the covering factors for the Lyman lines as
those derived from \ion{O}{6} fitting, then permit the covering factors to
vary only for the components whose fitting significantly deviates from the
observed absorption features after adjusting the parameters of column density.

Table~\ref{lyman} lists the results of fits for the Lyman series, giving
the column density, width, rest-frame velocity and covering factors in
Lyman lines through Ly$\epsilon$ for each kinematic component. Fig.~4 shows
the best fit to Ly$\alpha$ emission and the associated absorption lines.
For component A, the width (i.e., FWHM) is restricted to be 23 \kms by the
narrower absorption profile in \ion{O}{6} $\lambda$1038, and its column density
from the Lyman lines is also constrained by the corresponding unsaturated
absorption features in Ly$\delta$ and Ly$\epsilon$. This modeling fails to
produce a good fit for component A in Ly$\alpha$, but does fit the higher order
Lyman lines very well.  This might indicate that there exists additional low
column density or low-ionization gas with a larger velocity dispersion at the
same velocity, resulting in excess absorption seen only in Ly$\alpha$.

Generally speaking, to model the same strength of absorption feature,
absorbing gas with a smaller covering factor is required to have a higher
column density. The covering factor of component E is held to be 100\% for the
Lyman lines, since its \ion{H}{1} column density is strictly constrained by the
fact that its corresponding absorption feature cannot be found in the Ly$\beta$
or Ly$\delta$ lines.  There is a deep trough at component G in Ly$\gamma$,
even deeper than the corresponding troughs in Ly$\alpha$ and Ly$\beta$, so it
is unlikely to result from only the material responsible for component G in
Ly$\alpha$ and Ly$\beta$. This is likely to be an intergalactic Lyman$\alpha$
line at $z=0.0915$ messing up this trough, so we excluded component G in
Ly$\gamma$ from our fit.

In addition to the components found in the \ion{O}{6} and \ion{N}{5} lines,
components B and J are detected in the Lyman lines.  Only weak lower limits on
their covering factors can be estimated from the residual flux at the
absorption centers, and we assume their covering factors are 100\% in our
fitting.

\section{PHYSICAL CONDITIONS OF THE ABSORBERS}

It is well known that the ionization structure of photoionized gas is largely
dependent upon the strength and shape of the incident continuum over a
complete wavelength range, from the millimeter to X-ray bands. 3C~351 is a
radio-loud and soft X-ray weak quasar whose spectral energy distribution (SED)
is significantly different from the standard AGN continuum. Fortunately, a
great volume of observational data in a variety of wavebands has been
obtained for this quasar, which allows a more realistic modeling of the
photoionized gas.  Following \citet{mathur94}, we use the continuum data from
\citet{elvis94}. The presence of the ionized absorber leads to uncertainty in
determining the intrinsic X-ray spectral index $\alpha$ (for a power-law
$f_{\nu} \propto \nu^{-\alpha}$). The extrapolation of the IR continuum is
strongly limited by 1.3 mm observations \citep{antonucci88}, and the slope in
the submillimeter region is $-5/2$, similar to that of radio-quiet quasars. A
detailed comparison of the observed continuum with the so-called ``standard''
AGN continuum can be found in \citet{mathur94}.

To investigate the physical conditions of the intrinsic absorbers
with photoionization models, we used the latest version of Cloudy (C94)
\citep{fer98} in a similar way to \citet{brotherton01}. A hydrogen volume
density of $10^9$ cm$^{-3}$ and solar chemical abundance are assumed in our
modeling. Based on the observed column densities of \ion{H}{1}, \ion{N}{5} and
\ion{O}{6} for a specified kinematic component, we try to determine the total
hydrogen column ($N_H$) and the ionization parameter ($U$) which is defined as

\begin{eqnarray}
U = \frac{Q}{4\pi r^2 n_{_H} c},
\end{eqnarray}
where $Q$ is the total number of ionizing photons per second, and $n_{_H}$ is
the total number density of hydrogen. This method can lead to a unique
solution for the major components which have measurements of column
density ratios of both N(\ion{O}{6})/N(\ion{H}{1}) and
N(\ion{N}{5})/N(\ion{H}{1}), namely components A, C, D, E, F, G, H and I. For
components L and M which have only N(\ion{O}{6})/N(\ion{H}{1}) measurements,
the lack of a detectable \ion{N}{5} line provides an upper limit for the column
density, logN(\ion{N}{5})$<$14.0, which restricts double-valued solutions to
the lower ionization parameter.

Instead of directly creating a large grid of models, we used the OPTIMIZE
command in our computation to search for the model with the best fit of
observed columns in an efficient way. We tried to test the dependence
upon the hydrogen volume density $n_{_H}$,
and found that the modeling is not sensitive to the value of hydrogen
density at all. Physical parameters for some absorption components are shown in
Table~\ref{model}. Note that most components have relatively low total column
densities and ionization parameters.

\subsection{Broad Absorption Features}

The associated absorption features are dominated by the broad (FWHM $\geq$ 60 \kms)
components with centroids shifted roughly from $-$781 to $-$2206 \kms.
These components are marked above the spectrum in Fig. 3, including C, D, E,
F, G and I. It is clear that all these broad components are found to have
partial covering of the \ion{O}{6} line emission, which largely favors an
intrinsic origin for the associated absorption systems. Note that most
components are found to have ionization parameters of $0.13 < U < 0.32$,
and total hydrogen columns of order $10^{19}$ cm$^{-2}$. Compared with
the $ROSAT$ PSPC values for the X-ray warm absorbers, $N_H \sim 1.4\times
10^{22}$ cm$^{-2}$ and $U \sim 6.7$ \citep{mathur94}, the UV absorbing gas with
physical conditions inferred from the STIS spectrum seems {\em not} to
represent the same gas that is seen as the X-ray warm absorber.

In the STIS spectrum, component E has the widest velocity coverage (i.e., FWHM
$\sim$ 381 \kms) and largest column densities for the \ion{O}{6} and
\ion{N}{5} doublets. It is also found to have the largest column density ratio
of N(\ion{O}{6})/N(\ion{H}{1}), which yields the highest ionization
parameter of 0.32 and total hydrogen column of 1.55 $\times 10^{19}$ cm$^{-2}$.
Similarly, component D has a higher column density ratio of N(OVI)/N(HI), and
the resulting ionization parameter is also relatively high. Component G, with
a width of 99 \kms and a rest-frame velocity of $-$1699 \kms, is found to have
the largest column density for Lyman lines. This component has a comparatively
smaller ratio of N(\ion{N}{5})/N(\ion{H}{1}) and a higher total hydrogen column
density. Note that its covering factors are different for the high-ionization
doublets($\sim$ 80\%), Ly$\alpha$ ($\sim$65\%), and higher order Lyman lines
($\sim$100\%).

Component C is measured to have the largest column density ratio of
N(\ion{N}{5})/N(\ion{H}{1}), with a covering factor of $\sim$ 67\%, but its
physical condition is similar to components F, H, and I, with lower ionization
parameter and total hydrogen column density. In addition, component L
seems to be absent in \ion{N}{5} \wavwav{1238,1242}, and its column density
ratio of N(\ion{O}{6})/N(\ion{H}{1}) is smaller. The corresponding
absorber is modeled to have a lower ionization parameter of 0.095, under the
constraint of N(\ion{N}{5}) $<10^{14}$ cm$^{-2}$.

\subsection{Narrow Absorption Features}

Component A appears in the Lyman lines and both high-ionization doublets, with
the smallest outflow velocity, $\sim$ -37 \kms. The complete absorption at the
line cores in Ly$\alpha$ and \ion{O}{6} \wavwav{1032,1038} might indicate that
the absorber is located completely outside of the continuum and BELR of 3C~351.
The ISM of the host galaxy is likely to be responsible for this narrow
absorption component. However, the resulting lowest ionization parameter
of 0.069 and its small rest-frame velocity might suggest that it is associated
with the low-ionization gas in a putative rotating disk near the center of
3C~351. The absorption features in the Lyman lines at this component velocity
cannot be well fitted by a single component with the same width of $\sim$23
\kms as that of the \ion{O}{6} lines (see Fig. 4). The accretion disk might be
the host of the inferred additional low-ionization absorbing gas.

Components K and M are clearly shown in the \ion{O}{6} doublet,  and their
covering factors for the \ion{O}{6} lines are determined to be 45\% and 81\%,
respectively. The covering factor of component M for the Lyman lines is changed
to 47\%, and a lower ionization parameter is obtained with the constraint of
N(\ion{N}{5}) $<10^{14}$ cm$^{-2}$. Component K is strongly blended
with components M and L in Ly$\alpha$ and Ly$\beta$, and it is possible that
some of the \ion{H}{1} absorption that should have been assigned to component K
was inadvertently assigned to components L or M instead. Additionally,
components N and O may be present in Ly$\alpha$ line, but they fall in the
low sensitivity or gaps between orders. We exclude these three components in
our measurement. Taking the components B and J seen only in the Lyman lines
into account, there are five components (plus K, N and O) with unkonwn relative
ratios of N(\ion{O}{6})/N(\ion{H}{1}) and N(\ion{N}{5})/N(\ion{H}{1}).

\section{DISCUSSION}

Based on the {\em HST} STIS spectrum of 3C~351, we derived the column density,
covering factor, total hydrogen column and ionization parameter for each
kinematic component. These quantities can help us determine if the gas
responsible for the UV absorption features can also account for the X-ray
absorption. As one of the very few X-ray-quiet quasars with an X-ray
flux sufficient to obtain a spectrum with $ROSAT$, 3C~351 has been observed
twice with the PSPC \citep{nicastro99}, and an ionized absorber is present,
whose dominant feature is a strong \ion{O}{7}/\ion{O}{8} absorption edge. The
best fit to the PSPC spectrum taken on 1991 October, taking the preferred
folded model with a break at 0.37 keV, leads to an ionization parameter log$U$
of $0.78 - 1.08$ and a total column density $N_H$ of $1 - 2 \times 10^{22}$
cm$^{-2}$ \citep{mathur94}. This ionization parameter $U$ differs by a factor
of $\sim$ 30 from that inferred from the STIS far-UV spectrum, and the
difference in total column density is much larger. The multiple kinematic
components, low ionization parameters and total hydrogen column densities for
most of the UV absorption indicate that the UV and X-ray absorption is
unlikely to arise in the same material. Of course, there would be a
possibility that the X-ray absorber was much weaker when the STIS observations
were obtained, since the X-ray and STIS observations were taken years apart.

For the narrow absorption components B, J, K, N and O, for which the available
absorption species are limited, we still can model the physical conditions,
albeit with lower reliability. Component B is found to have only the low
ionization solution: ${\rm log} U \sim -1.24$ and ${\rm log} N_H \sim 18.11$.
However, two solutions, one with lower $U$ and $N_H$, and another with higher
$U$ and $N_H$, are found with acceptable $\chi^2$ values for components J,
K, N and O. For these four components, we put the additional limit of log
N(\ion{C}{3}) $<$ 12.5 in the modeling, considering the non-detection of the
\ion{C}{3} $\lambda$ 977 line, and two-fold solutions are still
obtained. Therefore, it is possible that a unified UV-X-ray absorber still
hides in these components. Components J and K are found to have similar higher
ionization solutions of log$U$ $\sim$ 1.25 and log$N_H$ $\sim$ 20.66, as well
as the lower ionization solutions: log$U$ $\sim$ -1.00 and log$N_H$ $\sim$
18.00 for component J; log$U$ $\sim$ 0.03 and log$N_H$ $\sim$ 18.02 for
component K. Components N and O seem to have similar high ionization solutions:
log $U \sim 1.29$ and log$N_H \sim 20.72$, and their low ionization solutions
show the low ionization parameters, log $U \sim -1.20$ for component N and
log$U \sim -1.17$ for component O. The physical conditions for these four
absorbers {\em may} be compatible with the X-ray warm absorber at this epoch,
so the possibility that some UV and X-ray absorption arises in the same gas
cannot yet be completely ruled out. Nevertheless our study suggests that the
absorbers associated with quasars are complex outflows and that the UV and
X-ray absorbing material are not identical for most individual kinematic
components. This picture is consistent with the model proposed by
\citet{krolik01}, for which material with a wide range of physical conditions
is stable.

As mentioned in \S 1, the radio properties of 3C~351 indicate an edge-on orientation
of the accretion disk, which seems to be consistent with the unified geometry of the
NAL and BAL absorbers detailed by Elvis (2000). According to this geometry,
we look across the flow direction in NALs, while we look down the length of the
flow in BALs (see Fig. 3 in Elvis 2000). The highest rest-frame velocity of the
outflowing ionized gas along the line of sight reachs -2800 \kms, so the
absorbing gas outflowing approximately perpendicular to the accretion disk can
be expected to have a much higher ejection velocity as shown in BALs ($0.1c -
0.2c$; Weymann et al. 1991).

There are techniques to derive additional information about the physical
characteristics of the UV absorber. The density and distance to the central
source of the associated absorber can be directly derived from time variability
of absorption lines \citep{hamann95}, but this technique can not be applied
for 3C~351 because it is a mildly variable quasar in the optical band, with a
factor of $\sim 2$ variability on time scales of $5 - 10$ yr \citep{grandi74}.
In the X-ray band, a factor of 1.7 change in flux within the nearly 2 years
between repeated $ROSAT$ PSPC observations is detected \citep{mathur94}. The
four runs of STIS observations for 3C~351 failed to reveal signification
UV variability over their time-frame, but some exploration is still possible.
From the definitions of column density and ionization parameter $U$ of the
absorber (Eq. 4), the thickness and the distance to the central engine can
be estimated from the gas density $n_{_H}$. If some absorbers lie outside of
the BELR, the constraints on the gas density can be used. As a result, we find
that the UV absorbers may be close to the central engine of $\sim$ 1 pc and a
typical thickness of $\sim 10^{12}$ cm.

It is interesting to point out that all the broad absorption components
(FWHM $>$ 60 \kms) appeared in both the \ion{O}{6} and \ion{N}{5} lines (i.e.,
components C, D, E, F, G, H and I) and have higher ionization parameters,
ranging from 0.13 to 0.25, than those of the narrow components. This might
indicate that the absorbing gas responsible for these broad absorption
components is closer to the central emission source. The turbulence in
rest-frame velocity might be an indicator of the location of the absorbing gas.
However, we can't find a similar tendency in the UV spectra of other AGNs.

Strong associated metal line absorption has been observed in AGNs with soft
X-ray absorption \citep{brandt00}. However, the disparity between the UV and
X-ray absorbing gas has been illustrated clearly in a handful of Seyfert
galaxies with both UV and X-ray observations. Only a small fraction have the UV
absorber identified with the X-ray absorber. For example, one of seven
components of intrinsic absorption in Seyfert 1 galaxy Mrk 509 is likely to be
associated with the warm X-ray-absorbing gas, and the associated X-ray
absorption for the other six components is negligible \citep{kriss00}. The
high-resolution X-ray spectrum for another well-studied Seyfert 1 galaxy, NGC
5548, taken with the Low Energy Transmission Grating (LETG) on $Chandra$, shows
strong narrow absorption lines with a blueshift of $\sim 280 \pm 70$ \kms,
which may be identified with the UV absorbing gas responsible for either
component 3 \citep{mathur99} or 4 \citep{brotherton01}. However,
high-resolution X-ray spectroscopy using $Chandra$\ has not yet been published
for 3C~351, so the kinematic link of the UV components with the X-ray absorbers
is not yet established.

We note the intriguing agreement in velocity among pairs of components in the
\ion{O}{6} doublet. The short-wavelength member of component A is nearly
centered in the long wavelength member of broad component G. Similarly, the
$\lambda$1038 lines of components N and O are superposed on the broad
$\lambda$1032 feature of component D. Line-locking is a possible cause of these
velocity alignments \citep{srianand00}, although chance superpositions in such
a complex configuration must also be probable.

Though multiple kinematic components have been resolved in this STIS spectrum,
the physical conditions of the absorbing gas are still poorly determined from
absorption lines studies in the UV band alone. To unambiguously identify the
UV absorption components with the X-ray warm absorbers, simultaneous
high-resolution spectroscopic observations are still needed for 3C~351. The lack
of X-ray spectroscopy of 3C~351 is currently a limit to better understanding the
relation between the warm absorbers and the UV associated absorbing gas.

\section{SUMMARY}

The $HST$ STIS spectroscopy of 3C~351 provides a good chance to investigate
the kinematic structure and physical conditions of the associated absorbers.
The high-resolution far-UV spectrum of the radio-loud and X-ray-weak quasar 3C
351 shows strong emission lines from the \ion{O}{6} \wavwav{1032,1038} and
Ly$\alpha$ lines, each of which can be resolved into two Gaussian profiles,
one broad and one narrow. Kinematically complex associated absorption in
3C~351 is present on the blue wings of the high-ionization emission doublets
\ion{O}{6} \wavwav{1032,1038}, \ion{N}{5} \wavwav{1238,1242} and Lyman lines
through Ly$\epsilon$. These absorption lines are resolved into
several distinct kinematic components, spanning a rest-frame velocity range
from $-$37 to $-$2800 \kms. All the broad absorption components are found to
have partial covering of the background continuum source and the BELR, which
strongly supports the intrinsic absorption origin and rules out the suggestin
that the absorption arises in some associated cluster of galaxies. The column
density ratios of N(\ion{O}{6})/N(\ion{H}{1}) and N(\ion{N}{5})/N(\ion{H}{1})
are used to derive the physical conditions for some major associated absorbers,
based on photoionization models. We find that all far-UV associated absorption
components have low values of the ionization parameter and effective hydrogen
column density, of which the higher ionization parameters are found for the
absorbers responsible for the broad (FWHM $>$ 60 \kms) absorption components.
Although the low spectral resolution in the X-ray band in previous studies
prevents a detailed comparison of the high-ionization resonance lines in the
far-UV and X-ray spectra, we still can conclude that the UV and X-ray
absorption are unlikely to arise in the same material, based on the multiple
kinematic components, low ionization parameters and total hydrogen column
densities we inferred for most of the UV absorbers.

\acknowledgments

The {\em STIS} observations were obtained for the STIS Instrument Definition
Team. QY, RFG, TMT, and MEK acknowledge NASA support for IDT analysis.
Additional support has been provided by NASA contract NAS5-30110. QY wants
to acknowledge the financial support provided by the China Scholarship Council
during the NOAO visit, the National Key Base Sciences Research Foundation under
contract G1999075402, and the Chinese NSF No.19833008, 19873018.

\clearpage

\begin{figure}
\epsscale{.7}
\plotone{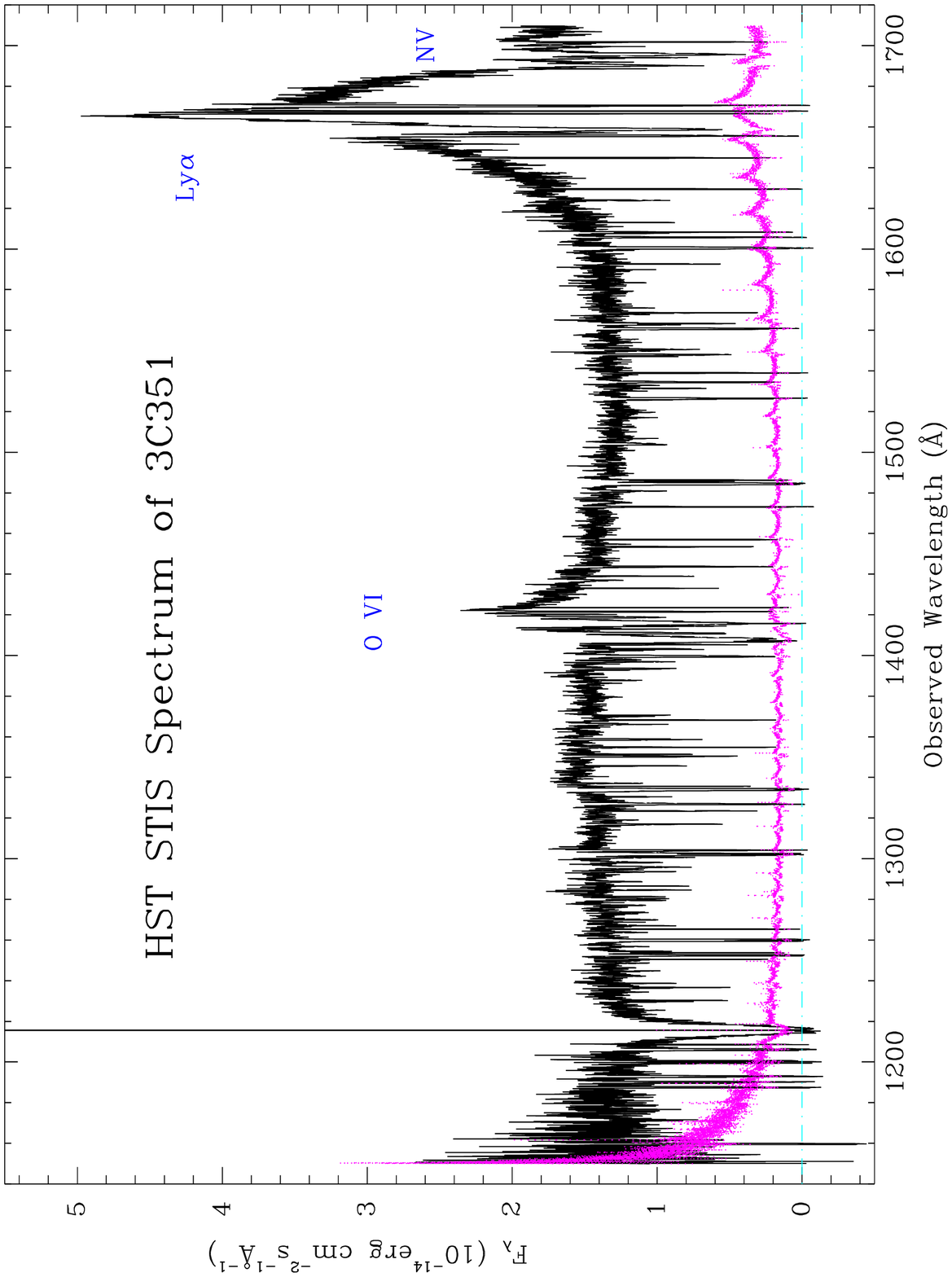}
\caption{$HST$ STIS far-UV spectrum of 3C~351,
smoothed by 9 pixels. 1$\sigma$ errors are shown as dotted lines. The spike at
1216 \AA\ is the geocoronal Ly$\alpha$ emission line, and the broad trough at
1216 \AA\ is the damped Ly$\alpha$ absorption line due to \ion{H}{1} in the
Milky Way ISM.}
\end{figure}

\clearpage

\begin{figure}
\epsscale{0.8}
\plotone{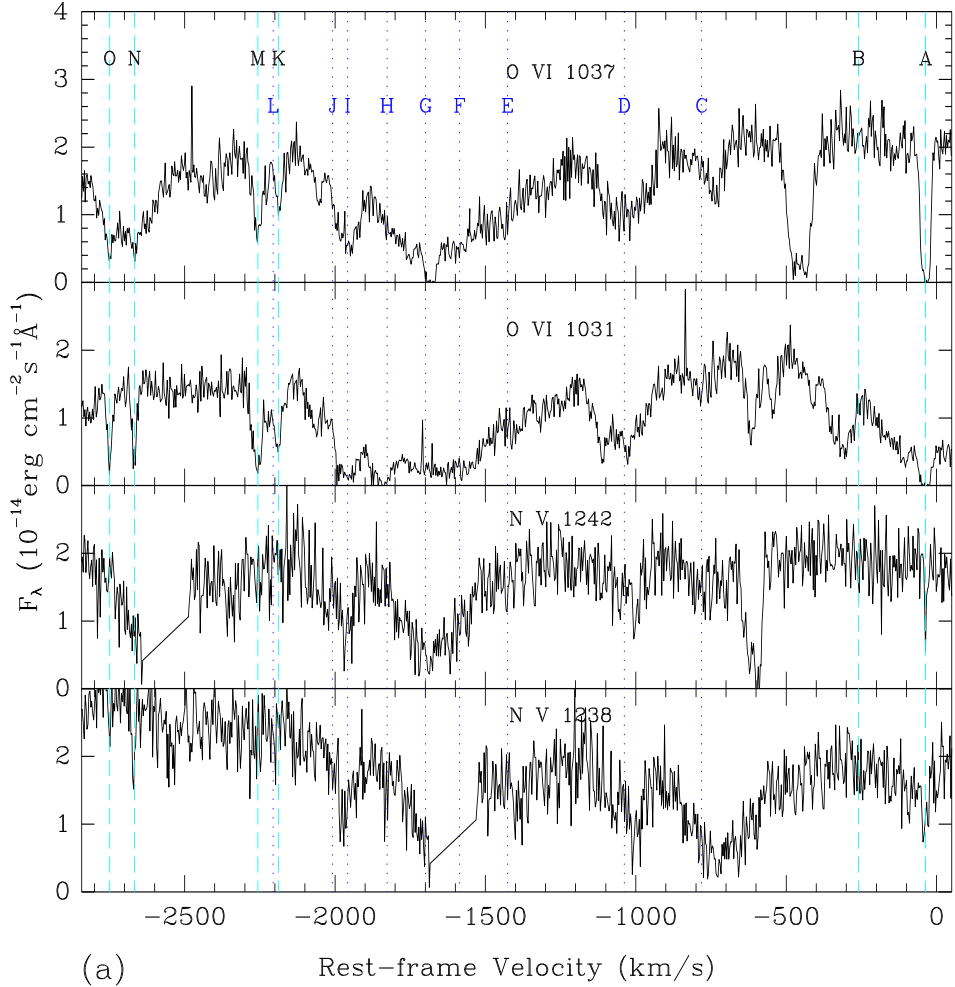}
\caption{The associated absorption features
in the rest-frame velocity domain. Fig. 2(a) is for the high-ionization
doublets \ion{O}{6} and \ion{N}{5}, and Fig. 2(b) for the Lyman series. The
vertical dotted lines represent the broad absorption components (FWHM $>$ 60
\kms) and the dashed lines stand for narrow components (FWHM $<$ 60 \kms). The
rest-frame velocity is relative to the systemic redshift of $z=0.3721$. The
covering factors for all these components can be well determined.}
\end{figure}

\clearpage
\begin{figure}
\epsscale{0.9}
\plotone{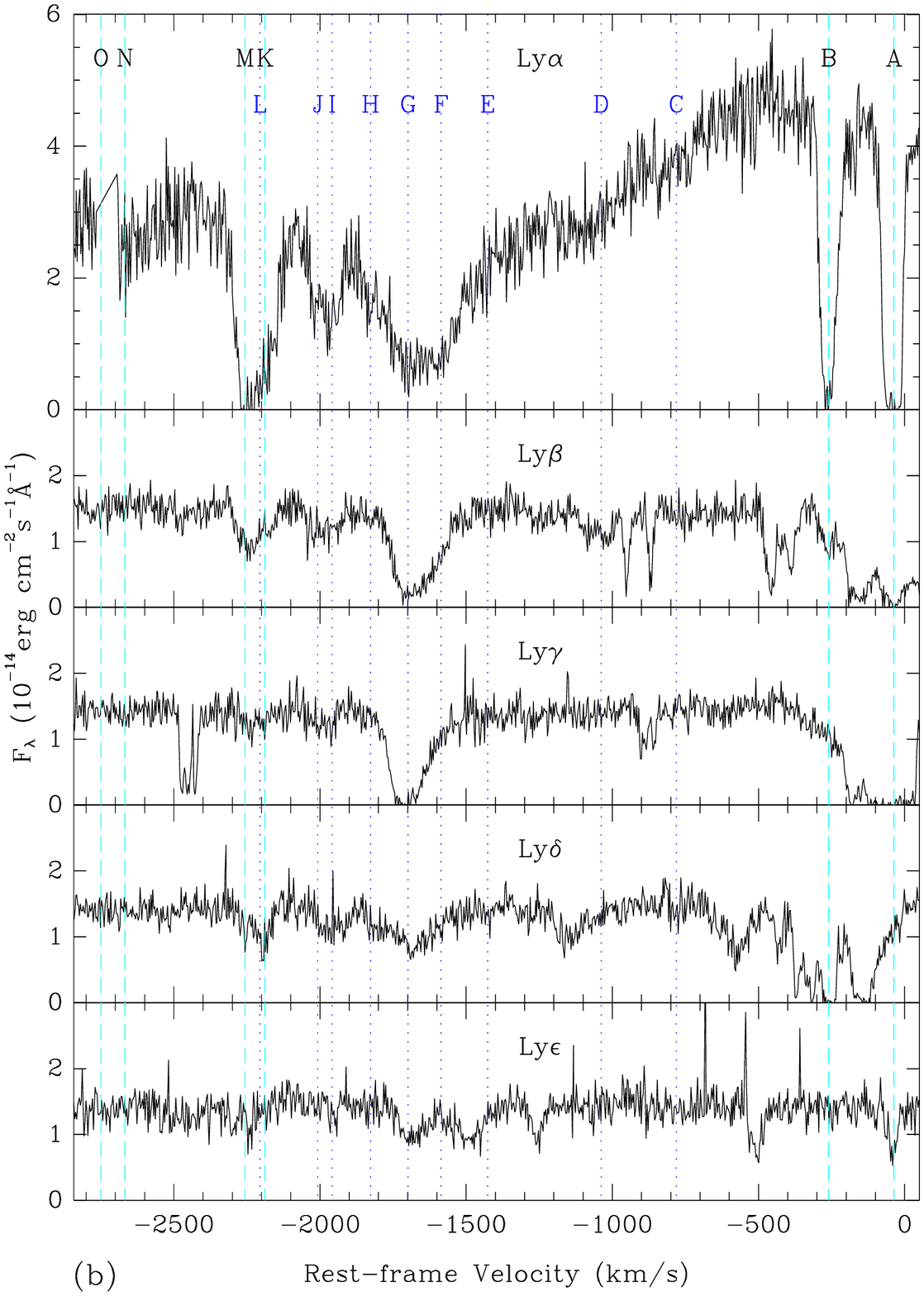}
\end{figure}

\clearpage

\begin{figure}
\epsscale{0.8}
\plotone{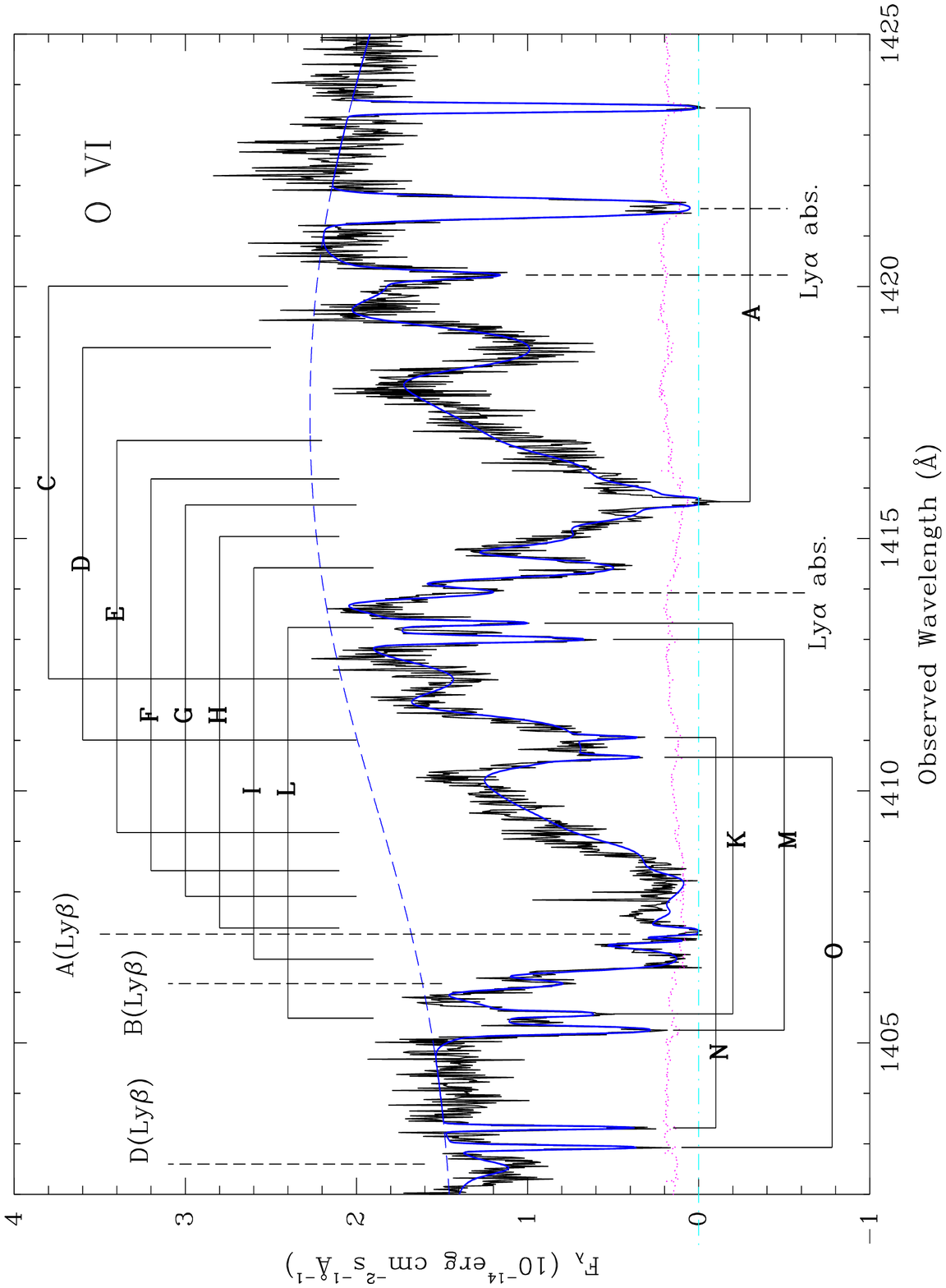}
\caption{Spectrum of emission of the
\ion{O}{6} doublet (\wavwav{1032,1038}) (dashed line) with the best fit (solid
line) to the associated absorption lines. The data are not binned, and
$1\sigma$ errors are shown with dotted lines. The broad absorption components
(FWHM $>$ 60 \kms) are marked with their designations above the spectrum, and
the narrow components (FWHM $<$ 60 \kms) are marked below the spectrum. Some
strong absorption features which do not belong to \ion{O}{6} are shown with
dashed lines. }
\end{figure}

\clearpage

\begin{figure}
\epsscale{0.8}
\plotone{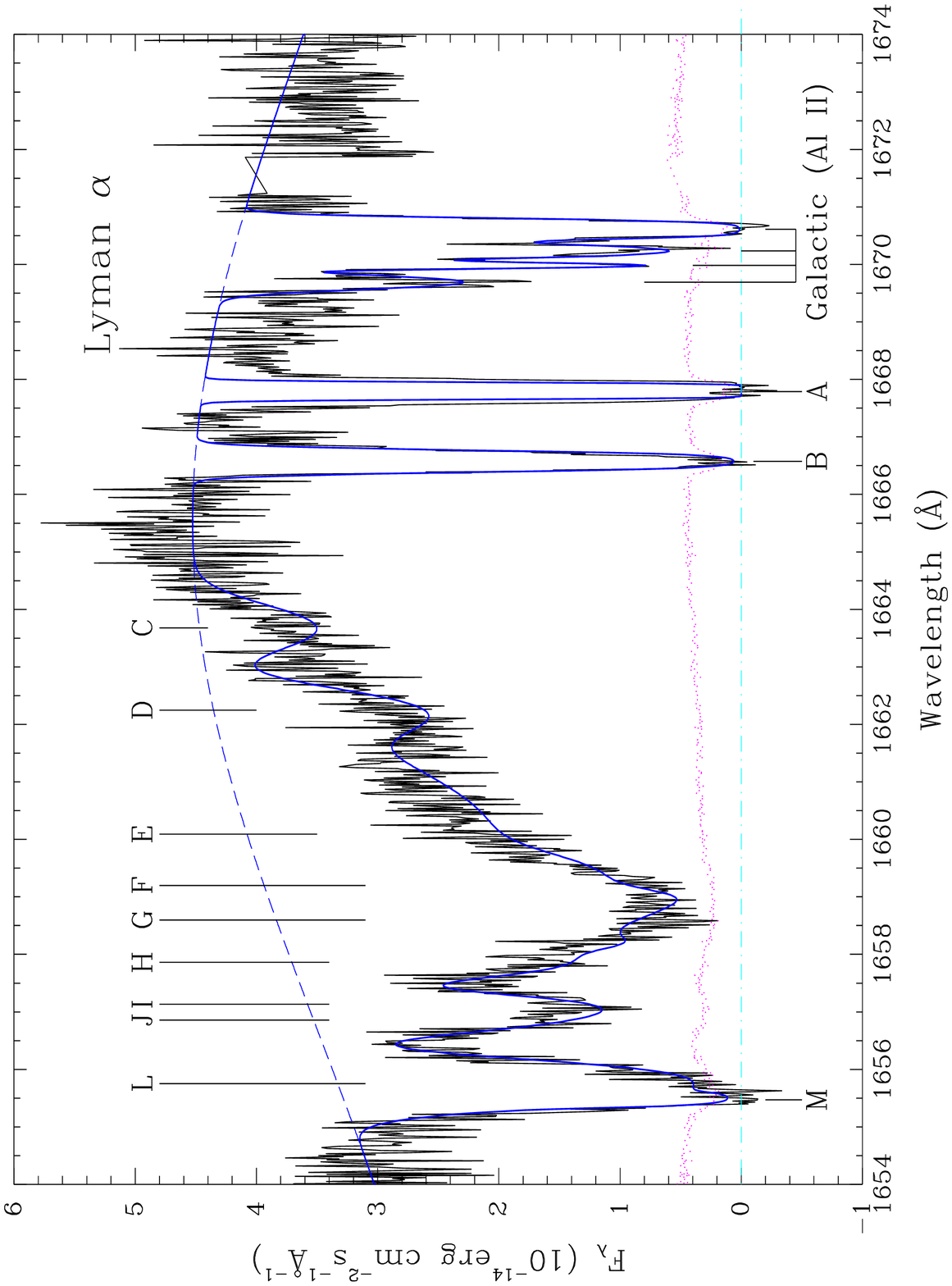}
\caption{Spectrum of Ly$\alpha$ emission
(dashed line) with the best fit (solid line) to the associated absorption
lines. The data are not binned, and $1\sigma$ errors are shown with dotted
lines. The broad absorption components (FWHM $>$ 60 \kms) are marked with
their designations above the spectrum, and the narrow components (FWHM $<$ 60
\kms) are marked below the spectrum. }
\end{figure}

\clearpage

\input{tab1.tex}

\end{document}

%% file: tab1.tex
\def\baselinestretch{1.2}

\begin{deluxetable}{lcc}
\tablewidth{0pc}
\tablecaption{Log of STIS Echelle Observations of 3C351\label{obslog}}
\tablehead{Observation & Total Integration& {\it HST} Archive\\
 Date & Time (seconds)& ID Codes}
\startdata
1999 June 27 & 19770 & O57901010--O57901080 \\
1999 June 29 & 14430 & O57903010--O57903060 \\
2000 Feb. 10 & 19770 & O57902010--O57902080 \\
2000 July 25 & 24228 & O57904010--O579040A0
\enddata
\end{deluxetable}


\clearpage

\begin{deluxetable}{clcll}
\tablecaption{ Emission lines in {\em HST} STIS far-UV spectrum of 3C~351
\label{emline}} \tablewidth{0pt}
\tablehead{ \colhead{Emission}  & \colhead{Observed $\lambda$ \tablenotemark{a}} &
\colhead{Flux\tablenotemark{b}} & \colhead{FWHM\tablenotemark{a}} & \colhead{Velocity} \\
\colhead{Lines} & \colhead{(\AA)} & \colhead{(10$^{-14}$ erg s$^{-1}$cm$^{-2}$)} &
\colhead{(\kms)} & \colhead{(\kms)} }
\startdata
Broad O {\small VI} $\lambda$1032 &  1423.64            & 11.0 $\pm$ 1.2 & 21041 $\pm$ 110 & 1639 \\
Broad O {\small VI} $\lambda$1038 &  1431.49            & 11.0 $\pm$ 1.7 & 21041           & 1639 \\
Broad Ly {$\alpha$} $\lambda$1216 &  1677.14 $\pm$ 0.02 & 85.5 $\pm$ 4.0 & 15001 $\pm$ 684 & 1639 $\pm$ 4 \\
Broad Ly {$\gamma$} $\lambda$~973 &  1341.70            & 15.0 $\pm$ 1.4 &
21041 & 1639 \\  \noalign{\smallskip} \noalign{\smallskip}
Narrow O {\small VI} $\lambda$1032 & 1413.56 $\pm$ 0.15 &  6.8 $\pm$ 0.5 & 2402 $\pm$  31 &  -450 $\pm$ 32 \\
Narrow O {\small VI} $\lambda$1038 & 1421.36            &  6.8 $\pm$ 0.3 & 2402           &  -450 \\
Narrow Ly {$\alpha$} $\lambda$1216 & 1665.26            & 49.8 $\pm$ 2.2 & 3601 $\pm$ 329 &  -450 \\
\enddata


\tablenotetext{a}{Values without uncertainties are either tied to another
          parameter or become fixed during the profile fitting.}
\tablenotetext{b}{Corrected for $E_{B-V} = 0.023$, $R_V=3.1$
          (Cardelli et al. 1989).}


\end{deluxetable}


\clearpage

\begin{deluxetable}{cccccl}
\tablecaption{Associated absorption lines
    of \ion{O}{6} and \ion{N}{5} doublets
\label{o6n5}}
\tablewidth{0pt}
\tablehead{
\colhead{Feature} & \colhead{Column density\tablenotemark{a}} &
\colhead{FWHM\tablenotemark{b}} & \colhead{Velocity\tablenotemark{b,c}} &
\colhead{Covering\tablenotemark{b}} \\
\colhead{} & \colhead{(10$^{12}$ cm$^{-2}$)} & 
\colhead{(\kms)} & \colhead{(\kms)} & \colhead{Fraction}
}
\startdata
   \multicolumn{2}{l}{\ion{O}{6} \wavwav{1032, 1038}}& & & & \\
   A&  490 ${\pm}$  66 &  23${\pm}$  1 &   -37 ${\pm}$ 4 & 1.00 \\
   C&  214 ${\pm}$  16 & 155${\pm}$ 24 &  -781 ${\pm}$16 & 0.67 ${\pm}$ 0.05 \\
   D& 1238 ${\pm}$ 107 & 138${\pm}$  8 & -1038 ${\pm}$ 9 & 0.62 ${\pm}$ 0.02 \\
   E& 3697 ${\pm}$ 271 & 381${\pm}$ 11 & -1426 ${\pm}$26 & 0.39 ${\pm}$ 0.02 \\
   F& 1186 ${\pm}$  94 & 156${\pm}$ 13 & -1586 ${\pm}$10 & 0.70 ${\pm}$ 0.01 \\
   G& 2238 ${\pm}$  59 &  99${\pm}$  4 & -1699 ${\pm}$ 5 & 0.80 ${\pm}$ 0.03 \\
   H&  626 ${\pm}$  80 & 101${\pm}$  2 & -1827 ${\pm}$ 6 & 0.85 ${\pm}$ 0.04 \\
   I&  541 ${\pm}$  38 &  75${\pm}$  4 & -1958 ${\pm}$ 5 & 0.96 ${\pm}$ 0.08 \\
   K&  185 ${\pm}$  75 &  14${\pm}$  1 & -2188 ${\pm}$ 5 & 0.45 ${\pm}$ 0.09 \\
   L&  314 ${\pm}$  53 & 116${\pm}$ 10 & -2206 ${\pm}$11 & 0.42 ${\pm}$ 0.01 \\
   M&  159 ${\pm}$ 148 &  23${\pm}$  6 & -2257 ${\pm}$ 6 & 0.81 ${\pm}$ 0.08 \\
   N&   46 ${\pm}$  15 &  14${\pm}$ 10 & -2667 ${\pm}$ 5 & 0.98 ${\pm}$ 0.13 \\
   O&   69 ${\pm}$  20 &  16${\pm}$  2 & -2749 ${\pm}$ 5 & 0.86 ${\pm}$ 0.21 \\
\noalign{\smallskip} \noalign{\smallskip}

  \multicolumn{2}{l}{\ion{N}{5} \wavwav{1238, 1242}} & & & & \\
 A &  25 ${\pm}$   5 &  23 &   -37 & 1.00 \\
 C & 239 ${\pm}$  67 & 155 &  -781 & 0.67 \\
 D & 134 ${\pm}$  42 & 138 & -1038 & 0.62 \\
 E & 708 ${\pm}$  22 & 381 & -1426 & 0.39 \\
 F & 395 ${\pm}$  42 & 156 & -1586 & 0.70 \\
 G & 367 ${\pm}$   8 &  99 & -1699 & 0.80 \\
 H &  59 ${\pm}$  23 & 101 & -1827 & 0.85 \\
 I &  81 ${\pm}$  24 &  75 & -1958 & 0.96 \\

\enddata

\tablenotetext{a}{Corrected for $E_{B-V} = 0.023$, $R_V=3.1$
          (Cardelli et al. 1989).}
\tablenotetext{b}{Values without uncertainties are either tied to another
          parameter or become fixed during profile fitting.}
\tablenotetext{c}{Velocities are relative to the systemic $z_{e}=0.3721 \pm
          0.0003$ ($2\sigma$) \citep{marziani96}. The uncertainty includes
          the spectral resolution and fitting error.}


\end{deluxetable}

\clearpage
\def\baselinestretch{1.0}
\begin{deluxetable}{ccccccccc}
\tablecaption{ Associated absorption lines
    of Lyman series \label{lyman}}
\tablewidth{0pt}
\tablehead{
\colhead{Feature} & \colhead{Column density \tablenotemark{a}} &
\colhead{FWHM} & \colhead{Velocity\tablenotemark{b,c}} & &
\colhead{Covering} & \colhead{factor\tablenotemark{b}}& & \\
\cline{5-9}
\colhead{} & \colhead{(10$^{12}$ cm$^{-2}$)} &\colhead{(\kms)} & \colhead{(\kms)}
& \colhead{Ly$\alpha$} &\colhead{Ly$\beta$} & \colhead{Ly$\gamma$} &
 \colhead{Ly$\delta$} & \colhead{Ly$\epsilon$}
}
\startdata
   A & 446 ${\pm}$  5 &  23            &   -37          & 1.00 &1.00 &1.00 &1.00 &1.00\\
   B & 106 ${\pm}$  5 &  43 ${\pm}$  3 &  -260 $\pm$  4 & 1.00 &1.00 &1.00 &1.00 & --\\
   C &  35 ${\pm}$  2 & 155            &  -781          & 0.67 & --  & --  & --  & --\\
   D &  55 ${\pm}$ 12 & 138            & -1038          & 0.62 &0.62 & --  & --  & --\\
   E & 144 ${\pm}$ 19 & 381            & -1426          & 1.00 & --  & --  & --  & --\\
   F & 147 ${\pm}$ 33 & 156            & -1586          & 0.70 &1.00 &1.00 & --  & --\\
   G & 629 ${\pm}$211 &  99            & -1699          & 0.65$\pm$0.06 &1.00 & --  &1.00 &1.00\\
   H &  70 ${\pm}$  7 & 101            & -1827          & 0.85 &0.85 &0.85 &0.85 & -- \\
   I &  35 ${\pm}$  5 &  75            & -1958          & 0.96 &0.96 &0.96 &0.96 & -- \\
   J &  36 ${\pm}$  6 &  90            & -2008 $\pm$ 49 & 1.00 &1.00 & -- & -- & -- \\
   L & 144 ${\pm}$ 32 & 116            & -2206          & 1.00 &1.00 & -- & -- & -- \\
   M & 124 ${\pm}$ 20 &  23            & -2257          & 0.47$\pm$0.31 &0.47 & -- & -- & -- \\

\enddata
\tablenotetext{a,b,c}{ Notes are given in Table~\ref{o6n5}}

\end{deluxetable}

\clearpage

\begin{deluxetable}{cccccc}
\tablecaption{Photoionization models for the major kinematic components
\label{model}}
\tablewidth{0pt}
\tablehead{
\colhead{Components} & \colhead{Velocity} &
\colhead{N(OVI)/N(HI)} &
\colhead{N(NV)/N(HI)} & \colhead{log $U$} &
\colhead{log $N_H$} \\
\colhead{} & \colhead{(\kms)}& \colhead{} & \colhead{} &
\colhead{} & \colhead{(cm$^{-2}$)}
}
\startdata
        A &   -37 &  1.10 &  0.06 & -1.16 & 18.36 \\
        C &  -781 &  6.11 &  6.83 & -0.86 & 18.37 \\
        D & -1038 & 22.51 &  2.44 & -0.51 & 18.65 \\
        E & -1426 & 25.67 &  4.92 & -0.50 & 19.19 \\
        F & -1586 &  8.07 &  2.69 & -0.76 & 18.86 \\
        G & -1699 &  3.56 &  0.58 & -0.91 & 19.10 \\
        H & -1827 &  8.94 &  0.84 & -0.71 & 18.39 \\
        I & -1958 & 15.46 &  2.31 & -0.63 & 18.36 \\
        L & -2206 &  2.18 &  --   & -1.02 & 18.43 \\
        M & -2257 &  1.28 &  --   & -1.13 & 18.30 \\
\enddata
\end{deluxetable}